\documentclass[conference,10pt]{IEEEtran}

\usepackage{cite}

\usepackage[utf8]{inputenc}
\usepackage{amsmath}
\usepackage{algorithm}
\usepackage{algorithmic}
\usepackage{xcolor}
\usepackage{amsmath, bm}
\usepackage{amsfonts, amssymb, amsbsy}

\usepackage{mathtools}

\usepackage{amsthm}

\makeatletter
\newcommand\fs@spaceruled{\def\@fs@cfont{\bfseries}\let\@fs@capt\floatc@ruled
	\def\@fs@pre{\vspace{0.5\baselineskip}\hrule height.8pt depth0pt \kern2pt}%
	\def\@fs@post{\kern1pt\hrule\relax}%
	\def\@fs@mid{\kern2pt\hrule\kern2pt}%
	\let\@fs@iftopcapt\iftrue}
\makeatother

\newtheorem{lemma}{Lemma}

\newtheorem{remark}{Remark}

\begin{document}
	\title{ Rate Balancing for Multiuser MIMO Systems}
	\author{
		\IEEEauthorblockN{Imène Ghamnia$^{1,2}$, Dirk Slock$^2$ and 	
			Yi Yuan-Wu$^{1}$\\ }
		\IEEEauthorblockA{
			\small
			$^{1}$Orange Labs, Châtillon, France,\\	
			$^{2}$Communication Systems Department, EURECOM, France \\		
			\{imene.ghamnia, yi.yuan\}@orange.com, dirk.slock@eurecom.fr	
			\vspace{-0.55cm}		}
		
	}
	\IEEEoverridecommandlockouts
	\maketitle
	\IEEEpeerreviewmaketitle
	\begin{abstract}
		We investigate and solve the rate balancing problem in the downlink for a multiuser Multiple-Input-Multiple-Output (MIMO) system. In particular, we adopt a transceiver structure to maximize the worst-case rate of the user while satisfying a total transmit power constraint. Most of the existing solutions either perform user Mean Squared Error (MSE) balancing or streamwise rate balancing, which is suboptimal in the MIMO case. The original rate balancing problem in the downlink is complicated due to the coupled structure of the transmit filters. This optimization problem is here solved in an alternating manner by exploiting weighted MSE uplink/downlink duality with proven convergence to a local optimum. Simulation results are provided to validate the proposed algorithm and demonstrate its performance improvement over unweighted MSE balancing.
	\end{abstract}
	\begin{IEEEkeywords}
		rate balancing, max-min fairness, MSE duality, tranceiver optimization, multiuser MIMO systems
		\vspace{-2mm}
	\end{IEEEkeywords}
	
	\section{Introduction}
	\label{sec:introduction}
	\vspace{-2mm}
	
	One important criterion in designing wireless networks is ensuring faireness requirements. Fairness is said to be achieved if some performance metric is equally reached by all users of the system, depending on their priority allocations. With   respect to applications in communication networks, fairness is closely related to min-max or max-min optimization problems, also referred to as \textit{balancing} problems. Actually, balancing a given metric or a utility function among users implies that the system performances are limited by the weak users. At the optimum, the performance of the latter is brought to be improved \cite{Zheng18}.

	However, most of balancing optimization problems are non-convex and can not be solved directly. Despite that, several works over the litterature have developped optimal solutions. For instance, \cite{Wiesel06} solved the max-min problem by a sequence of Second Order Cone Programs (SOCP). Also, \cite{Bengtsson1} showed that a semidefinite relaxation is tight for the problem, and the optimal solution can be constructed from the solution to a reformulated semidefinite program. In \cite{Cai11}, the authors proposed an algorithm based on fixed-point that alternates between power update and beamformer updates, and the nonlinear Perron-Frobenius theory was applied to prove the convergence of the algorithm.
	
	Another way to solve balancing optimization problems  is to convert the problem from the downlink (DL) channel to its equivalent uplink (UL) channel, by exploiting the UL/DL \textit{duality}. Doing so, the transformed problem has better mathematical structure and convexity in the UL, thus, the computational complexity of the original problem can be reduced \cite{Boche4}. The UL/DL duality has been widely used to design optimal transmit and receive filters that ensure faireness requirements w.r.t.\  the Signal-to-Interference-plus-Noise Ratio (SINR), the Mean Square Error (MSE), and the user or stream rate.
	
	
	
	%
	With the objective being to equalize all user SINRs, the SINR balancing problem is of particular interest because it is directly related to common performance measures like system capacity and bit error rates. Maximizing the minimum user SINR in the UL can be done straightforwardly since the beamformers can be optimized individually and SINRs are only coupled by the users’ transmit powers. In contrast, DL optimization is generally a nontrivial task because the user SINRs depend on all optimization variables and have to be optimized jointly \cite{Mont98,Negro11, Boche4,Yu7,Zhang9,Cumanan10}. 
	
	%
	Another well-known duality is the stream-wise MSE duality where it has been shown that the same MSE values are achievable in the DL and the UL with the same transmit power constraint. This MSE duality has been exploited to solve various minimum MSE (MMSE) based optimization problems \cite{SSB,shi7,hunger9}. 
	%
	%
	
	In this work, we focus on user rate balancing in a way to maximize the minimum (weighted) rate among all the users in the cell, in order to achieve cell-wide fairness. This balancing problem is studied in \cite{111}. However, the authors do not provide an explicit precoder design. Here we
	provide a solution via the relation between user rate (summed
	over its streams) and a weighted sum MSE. But also another ingredient is required: the exploitation of a scale factor that can be freely chosen in the weights for the weighted rate balancing.
	User-wise rate
	balancing outperforms user-wise MSE balancing or streamwise
	rate (or MSE or SINR) balancing when the streams of
	any MIMO user are quite unbalanced. 
	\section{System Model}
	\label{sec:sys}
	
	\begin{figure}
		\centering
		\includegraphics[width=250pt]{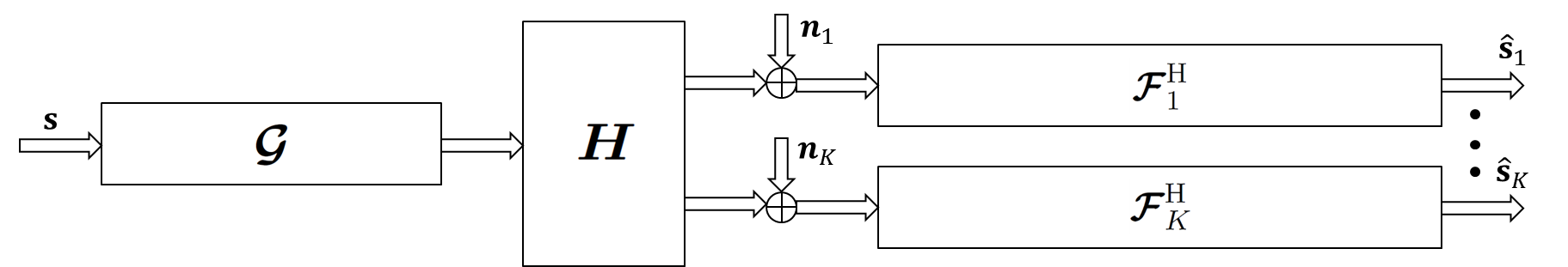}
		{\small (a)}
		\includegraphics[width=250pt]{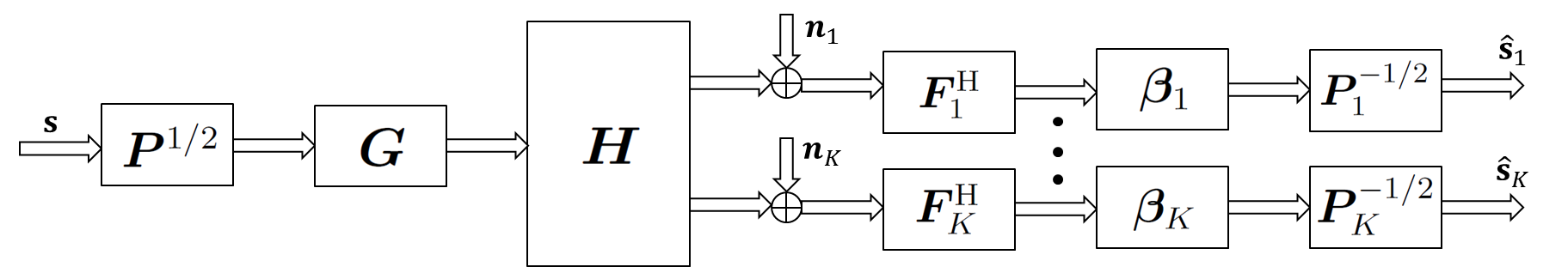}
		{\small (b) }
		\vspace{-2mm}
		\caption{System model: (a) DL channel, (b) equivalent DL channel.}
		\label{fig:dl}
		\vspace{-4mm}
	\end{figure}
	
	The considered network is a multiuser MIMO DL system, (see Figure \ref{fig:dl}). We focus on a Base Station (BS) of $ M $ transmit antennas serving $K$ users of each $N_k$ antennas, ($k = 1, ..., K$ is the users' index). The channel between the $k$th user and the BS is denoted by $ \bm{H}_{k}^{\mathrm{H}} \in \mathbb{C}^{M \times N_k}$, and $\bm{H}^{\mathrm{H}}= [\bm{H}_{1}^{\mathrm{H}}, ..., \bm{H}_{K}^{\mathrm{H}}]$ is the overall channel matrix.
	
	We assume zero-mean white Gaussian noise $ \bm{n}_k \in \mathbb{C}^{N_k \times 1} $ with distribution $ \mathcal{CN}(0, \sigma_{n}^{2}\bm{I})$ at the $k$th user. We assume independent unity-power transmit symbols $ {\bf s} = [{\bf s}_{1}^{\mathrm{T}} \ldots {\bf s}_{K}^{\mathrm{T}} ]^{\mathrm{T}}$, i.e., $\mathbb{E} \big[ {\bf s} {\bf s}^{\mathrm{H}}\big] = \bm{I}  $, where ${\bf s}_k \in \mathbb{C}^{d_k \times 1} $ is the data vector to be transmitted to the $k$th user, with $d_k$ being the number of streams allowed by user $k$. The latter are transmitted using the transmit filtering matrix  $\bm{\mathcal{G}} = \bm{G} \bm{P}^{1/2}\in \mathbb{C}^{M \times N_d}  $, composed of the beamforming matrix $ \bm{G} = [ \bm{G}_{1} \ldots \bm{G}_{K}] = [ \bm{g}_{1} \ldots \bm{g}_{N_d}] $ with normalized columns $\lVert \bm{g}_{i} \rVert_{2} = 1 $ and the diagonal non-negative DL power allocation $\bm{P}^{1/2} = \text{blkdiag} \{ \bm{P}_{1}^{1/2}, \ldots, \bm{P}_{K}^{1/2} \} $ where $ \text{diag} ( \bm{P}_k) \in \mathbb{R}_{+}^{d_k \times 1} $ contains the transmission powers and $ N_d = \sum_{k=1}^{K} d_k $ is the total number of streams. The total transmit power is limitted, i.e., $ \mathrm{tr} \big(  \bm{P})  \leq P_{\max}$. 
	
	Similarly, the receive filtering matrix for each user is defined as $\bm{\mathcal{F}}_{k}^{\mathrm{H}} = \bm{P}_{k}^{-1/2} \bm{\beta}_k \bm{F}_{k}^{\mathrm{H}} \in \mathbb{C}^{d_k \times N_k} $, composed of beamforming matrix $\bm{F}_{k}^{\mathrm{H}} \in \mathbb{C}^{d_k \times N_k} $ and the diagonal matrices $ \bm{\beta}_{k}$ contain scaling factors which ensure that the columns of $ \bm{F}_{k}^{\mathrm{H}}$ have unit norm. We define $ \bm{\beta} = \text{blkdiag} \{ \bm{\beta}_{1}, \ldots, \bm{\beta}_{K}\} = \mathrm{diag}\{ [ \beta_{1} \ldots \beta_{N_d}] \}$ and $ \bm{F} = \text{blkdiag} \{ \bm{F}_{1}, \ldots, \bm{F}_{K} \} = [ \bm{f}_{1} \ldots \bm{f}_{N_d}] $ with normalized per-stream receivers, i.e., $\lVert \bm{f}_{i} \rVert_{2} = 1 $.
	
	The MSE per stream $ \varepsilon_{i}^{\mathrm{DL}} $ between the decision variable $\hat{s}_i $ and the transmit data symbol $s_i$ is defined as follows
	\vspace{-3mm}
	\begin{small}
		\begin{align} \label{eq:dlmse}
		\varepsilon_{i}^{\mathrm{DL}} &= \mathbb{E} \Big\{ \lvert \hat{s}_i - s_i \rvert^{2} \Big\}
		= \beta_{i}^{2}/p_i \bm{f}_{i}^{\mathrm{H}} \bm{H} \big(\sum_{j=1}^{N_d} p_j \bm{g}_j\bm{g}_{j}^{\mathrm{H}} \big) \bm{H}^{\mathrm{H}} \bm{f}_{i} 
		\nonumber \\[-2mm]
		&- 2 \beta_{i} \mathrm{Re} \big\{ \bm{f}_{i}^{\mathrm{H}} \bm{H} \bm{g}_i\big\} + \sigma_{n}^{2}\beta_{i}^{2}/ p_i +1, \forall i \in \{ 1,...,N_d \}. 
		\end{align}
	\end{small}
	\vspace{-5mm}
	
	\section{Problem Formulation} \label{sec:pb}
	
	In this work, we aim to solve the weighted user-rate max-min optimization problem under a total transmit power constraint, i.e., the user rate balancing problem expressed as follows
	\vspace{-4mm}
	\begin{small}
		\begin{align}  \label{eq:main}
		\max_{\{\bm{G}, \bm{P}, \bm{F}, \bm{\beta} \}} \min_{k} &  \; \; r_{k}/ r_{k}^{\circ} \; \; \nonumber
		\\ \mathrm{s.t.} & \; \; \mathrm{tr}\big(\bm{P}) \leq P_{\max} 
		\vspace{-3mm}
		\end{align}
	\end{small}
	where $r_{k} $ is the $k$th user-rate
	\begin{small}
		\begin{equation} \label{eq:rate}
		r_{k} = \ln \mathrm{det} \Big( \bm{I} \!+\! \bm{H}_{k} \bm{\mathcal{G}}_{k} \bm{\mathcal{G}}_{k}^{\mathrm{H}} \bm{H}_{k}^{\mathrm{H}} \big( \sigma_{n}^{2} \bm{I}\!+\! \sum_{j \neq k} \bm{H}_{k} \bm{\mathcal{G}}_{j} \bm{\mathcal{G}}_{j}^{\mathrm{H}} \bm{H}_{k}^{\mathrm{H}}\big)^{-1} \Big)
		\end{equation} 
	\end{small}
	and $r_{k}^{\circ}$ is the rate scaling factor for user $k$.
	However, the problem presented in (\ref{eq:main}) is complex and can not be solved directly. 
	\begin{lemma} \label{lem:rate}
		The rate of user $k$ in (\ref{eq:rate}) can also be represented as
		\vspace{-7mm}
		\begin{align} \label{eq:lem}
		r_{k} =   \max_{\bm{W}_{k}, \bm{\mathcal{F}}_{k}} \big[ \ln \mathrm{det} \big( \bm{W}_{k} \big) - \mathrm{tr} \big( \bm{W}_{k}  \bm{\mathrm{E}}_{k}^{\mathrm{DL}}  \big) + d_k \big].
		\vspace{-4mm}
		\end{align}
		where \vspace{-6mm}
		\begin{small}
			\begin{align} \label{eq:5}
			\bm{\mathrm{E}}_{k}^{\mathrm{DL}} &= \mathbb{E} \Big[ ( \hat{{\bf s}}_k - {\bf s}_k )( \hat{{\bf s}}_k - {\bf s}_k )^{\mathrm{H}} \Big] \nonumber
			\\
			&= (\bm{I} - \bm{\mathcal{F}}_{k}^{\mathrm{H}} \bm{H}_{k} \bm{\mathcal{G}}_{k})(\bm{I} - \bm{\mathcal{F}}_{k}^{\mathrm{H}} \bm{H}_{k} \bm{\mathcal{G}}_{k})^{\mathrm{H}} \nonumber \\&+ \sum_{j \neq k} \bm{\mathcal{F}}_{k}^{\mathrm{H}} \bm{H}_{j} \bm{\mathcal{G}}_{j}\bm{\mathcal{G}}_{j}^{\mathrm{H}} \bm{H}_{j}^{\mathrm{H}} \bm{\mathcal{F}}_{k} + \sigma_{n}^{2} \bm{\mathcal{F}}_{k}^{\mathrm{H}}\bm{\mathcal{F}}_{k}	
			\end{align}
		\end{small}
		is the $k$th-user DL MSE matrix between the decision variable $ \hat{{\bf s}}_k  $ and the transmit signal ${ \bf s}_k  $, and $\bm{W} = \{ \bm{W}_k \}_{1 \leq k \leq K} $ are auxiliary weight matrix variables with optimal solution $\bm{W}_{k} = \big ( \bm{\mathrm{E}}_{k}^{\mathrm{DL}}\big )^{-1} $ and $ \bm{\mathcal{F}}_{k} = (\sigma_{n}^{2} \bm{I} + \sum_{j= 1}^{K} \bm{H}_{k} \bm{\mathcal{G}}_{j}\bm{\mathcal{G}}_{j}^{\mathrm{H}}\bm{H}_{k}^{\mathrm{H}})^{-1} \bm{H}_{k}\bm{\mathcal{G}}_{k} $,  \cite{Razaviyayn11}.
	\end{lemma}
	%
	
	Now considering both (\ref{eq:main}) and (\ref{eq:lem}), and introducing
	$t = \min_{k} r_{k}/ r_{k}^{\circ} $, we have $\forall k  $
	\vspace{-2mm}
	\begin{small}
		\begin{align}
		r_{k}/ (t\, r_{k}^{\circ})\geq 1 \mbox{ or }
		r_{k}/ r_{k}^{\circ} &\geq t \nonumber\\
		\stackrel{(a)}{\Longleftrightarrow}	\ln \mathrm{det} \big( \bm{W}_{k} \big) +  d_k -\mathrm{tr} \big( \bm{W}_{k}  \bm{\mathrm{E}}_{k}^{\mathrm{DL}}  \big) &\geq t r_{k}^{\circ} \label{eq:equiv}\\
		\Longleftrightarrow\frac{ \mathrm{tr} \big( \bm{W}_{k}  \bm{\mathrm{E}}_{k}^{\mathrm{DL}}  \big)}{\ln \mathrm{det} \big( \bm{W}_{k} \big) +  d_k - tr_{k}^{\circ}} \stackrel{(b)}{=} \frac{\epsilon_{w,k}^{\mathrm{DL}}}{\xi_{k}} &\leq 1 \nonumber 
		\end{align}
	\end{small}
	where $(a)$ follows from (\ref{eq:lem}) (with optimal $\bm{W}_{k}$) and $(b)$ from  $\epsilon_{w,k}^{\mathrm{DL}} = \mathrm{tr} \big( \bm{W}_{k}  \bm{\mathrm{E}}_{k}^{\mathrm{DL}}  \big)$, the matrix-weighted MSE (WMSE), and $ \xi_{k}= \ln \mathrm{det} \big( \bm{W}_{k} \big) + d_k-r_{k}^{\vartriangle}$ the WMSE requirement, with $ r_{k}^{\vartriangle}=t r_{k}^{\circ}$ the individual rate target, i.e. $r_{k} \geq r_{k}^{\vartriangle}  $. 
	What we exploit here is a scale factor $t$ that can be chosen freely in the rate weights $r^o_k$ in (\ref{eq:main}),
	to transform the rate weights $r^{\circ}_k$ into target rates $r^{\vartriangle}_k = t r^{\circ}_k$, which at the same time
	allows to interpret the WMSE weights $\xi_k$ as target WMSE values.
	
	Doing so, the initial rate balancing optimization problem (\ref{eq:main}) can be transformed into a matrix-weighted MSE balancing problem expressed as follows
	\begin{small}
		\begin{align}  \label{eq:primal}
		\min_{\{ \bm{G}, \bm{P}, \bm{F}, \bm{\beta} \}} \max_{k} &  \; \; \epsilon_{w,k}^{\mathrm{DL}} / \xi_{k} \; \; \nonumber
		\\ \mathrm{s.t.} & \; \; \mathrm{tr}\big(\bm{P} \big)  \leq P_{\max}, 
		\end{align}
	\end{small}
	which needs to be complemented with an outer loop in which $\bm{W}_{k} = \big( \bm{\mathrm{E}}_{k}^{\mathrm{DL}}\big )^{-1} $,
	$t = \min_{k} r_{k}/ r_{k}^{\circ} $, $r^{\vartriangle}_k = t r^{\circ}_k$
	and  $ \xi_{k}=  d_k + r_k -r_{k}^{\vartriangle}$ get updated.

	The problem in (\ref{eq:primal}) is still difficult to be handled directly. In the next sections, we solve the problem via UL and DL MSE duality. To this aim, we model an equivalent UL-DL channel plus transceivers pair by separating the filters into two parts: a matrix with unity-norm columns and a scaling matrix \cite{shi8}. Then, the UL and DL are proved to share the same MSE by switching the role of the normalized filters in the UL and DL. Doing so, an algorithmic solution can be derived for the optimization problem (\ref{eq:primal}).
	\vspace{-2mm}
	\section{Dual UL Channel}
	\label{sec:eq}
	
	\begin{figure}
		\centering
		\includegraphics[width=250pt]{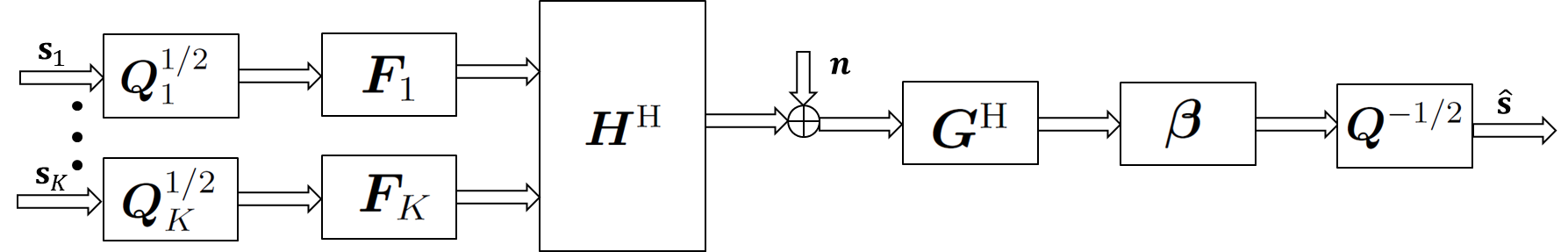}
		\caption{Dual UL channel.}
		\vspace{-4mm}
		\label{fig:dualup}
	\end{figure}
	In the equivalent UL model represented in Figure \ref{fig:dualup}, we switch between the role of the normalized transmit and receive filters. In fact, $ \bm{F}_{k} \bm{Q}_{k}^{1/2}$ is the $k$th transmit filter and $ \bm{Q}^{-1/2} \bm{\beta} \bm{G}^{\mathrm{H}} $ is a multiuser receive filter, where $\bm{Q} = \text{blkdiag}\big\{ \bm{Q}_1, ..., \bm{Q}_K\big\} $ with $ \text{diag} ( \bm{Q}_k) \in \mathbb{R}_{+}^{d_k \times 1} $ being the UL power allocation.
	
	Although the quantities $\bm{H}, \bm{G}, \bm{F}$ and $ \bm{\beta}$ are the same, the UL power allocation $\bm{q} = [q_1 \ldots q_{N_d}]^{\mathrm{T}}  = \text{diag} ( \bm{Q}) $ may differ from the DL allocation $\bm{p} = [p_1 \ldots p_{N_d}]^{\mathrm{T}}  = \text{diag} ( \bm{P}) $, both verifying the same sum power constraint $ \lVert \bm{p}\rVert_{1} =   \lVert \bm{q}\rVert_{1} \leq P_{\max}$. 
	
	The corresponding UL per stream MSE  $ \varepsilon_{i}^{\mathrm{UL}} $ is given by
	\begin{small}
		\begin{align} \label{eq:upmse}
		\varepsilon_{i}^{\mathrm{UL}} &= \beta_{i}^{2}/q_i \bm{g}_{i}^{\mathrm{H}} \bm{H}^{\mathrm{H}} \big(\sum_{j=1}^{N_d} q_j \bm{f}_j\bm{f}_{j}^{\mathrm{H}} \big) \bm{H} \bm{g}_{i} - 2 \beta_{i} \mathrm{Re} \big\{ \bm{g}_{i}^{\mathrm{H}} \bm{H}^{\mathrm{H}} \bm{f}_i\big\} \nonumber \\ & + \sigma_{n}^{2}\beta_{i}^{2}/ q_i +1, \forall i .
		\end{align}
	\end{small}
	\vspace{-6mm}
	
	\section{MSE Duality}
	\label{sec:dual}
	With the equivalent DL channel and its dual UL, it has been shown that the same per stream MSE values are achieved in both links, i.e.,  $ \bm{\varepsilon}^{\mathrm{UL/DL}} = \text{diag}\big\{ [ \varepsilon_{1}^{\mathrm{UL/DL}} \ldots \varepsilon_{N_d}^{\mathrm{UL/DL}}]\big\}  = \text{diag}\big\{ [ \varepsilon_{1} \ldots \varepsilon_{N_d}]\big\} = \bm{\varepsilon}$ \cite{shi8}.
	\vspace{1mm}
	
	The UL and DL power allocation, obtained by solving the MSE expressions as in (\ref{eq:upmse}) for UL w.r.t. the powers, are given by
	\vspace{-2mm}
	\begin{small}
		\begin{equation} \label{eq:q}
		\bm{q} = \sigma_{n}^{2} (\bm{\varepsilon} - \bm{D} - \bm{\beta}^{2} \bm{\Psi})^{-1} \bm{\beta}^{2} \bm{1}_{N_d}
		\vspace{-2mm}
		\end{equation}
	\end{small}
	and\vspace{-2mm}
	\begin{small}
		\begin{equation} \label{eq:p}
		\bm{p} = \sigma_{n}^{2} (\bm{\varepsilon} - \bm{D} - \bm{\beta}^{2} \bm{\Psi}^{T})^{-1} \bm{\beta}^{2} \bm{1}_{N_d}
		\end{equation}
	\end{small}
	respectively, where the diagonal matrix $\bm{D}$ is defined as 
	\begin{small}
		$$[\bm{D}]_{ii} = \beta_{i}^{2} \bm{g}_{i}^{\mathrm{H}} \bm{H}^{\mathrm{H}}\bm{f}_i \bm{f}_{i}^{\mathrm{H}} \bm{H} \bm{g}_i - 2 \beta_{i} \mathrm{Re} \{\bm{g}_{i}^{\mathrm{H}} \bm{H}^{\mathrm{H}} \bm{f}_i \} + 1
		\vspace{-2mm}$$
	\end{small}
	and\vspace{-2mm}
	\begin{small}
		\begin{equation}
		[\bm{\Psi}]_{ij} = \begin{cases}
		\bm{g}_{i}^{\mathrm{H}} \bm{H}^{\mathrm{H}} \bm{f}_j \bm{f}_{j}^{\mathrm{H}} \bm{H} \bm{g}_i,
		& i \neq j \\
		0,			
		& i=j.
		\end{cases}	\nonumber
		\end{equation}
	\end{small}
	In fact, the MSE duality allows to optimize the transceiver design by switching between the virtual UL and actual DL channels. The optimal receive filtering matrices in both UL and DL are MMSE filters and given by
	\begin{small}
		\vspace{-1mm}
		\begin{equation} \label{rup}
		\bm{G}_k \bm{\beta}_k \bm{Q}_{k}^{-1/2} = \big( \bm{H}^{\mathrm{H}} \bm{ F Q F}^{\mathrm{H}} \bm{H} + \sigma_{n}^{2} \bm{I} \big)^{-1} \bm{H}_{k}^{\mathrm{H}} \bm{F}_k \bm{Q}_{k}^{1/2}
		\end{equation}
	\end{small}
	and
	\begin{small}
		\begin{equation} \label{rdl}
		\bm{F}_k \bm{\beta}_k \bm{P}_{k}^{-1/2} = \big( \bm{H}_{k} \bm{G P G}^{\mathrm{H}} \bm{H}_{k}^{\mathrm{H}}+ \sigma_{n}^{2} \bm{I} \big)^{-1} \bm{H}_{k} \bm{G}_k \bm{P}_{k}^{1/2}.
		\end{equation}
		\vspace{-1mm}
	\end{small}
	\vspace{-5mm}
	
	\section{The matrix weighted User-MSE Optimization}
	\label{sec:sol}
	
	In this section, the problem (\ref{eq:primal}) with respect to the  matrix weighted user-MSE is studied. First, we start by the UL power allocation strategies. Then, the joint optimization will follow given the MSE duality.
	In fact, the MSE duality opens  up  a  way  to  obtain  optimal  MMSE receiver designs in (\ref{rup}) and (\ref{rdl}). The DL matrix weighted user-MSE optimization problems can be solved by optimizing the weighted MSE values of the dual UL system. 
	The latter can be formulated as
	\begin{small}
		\vspace{-4mm}
		\begin{align}  \label{eq:primalup}
		\min_{\{ \bm{G}, \bm{F}, \bm{W} \}} \max_{k} &  \; \; \epsilon_{w,k}^{\mathrm{UL}} / \xi_{k} \; \; \nonumber
		\\ \mathrm{s.t.} & \; \; \mathrm{tr}\big(\bm{Q}\big)  \leq P_{\max}
		\end{align}
	\end{small}
	where $\epsilon_{w,k}^{\mathrm{UL}} = \mathrm{tr} \big( \bm{W}_{k}  \bm{\mathrm{E}}_{k}^{\mathrm{UL}}  \big)$, and
	\begin{small}
		\vspace{-1mm}
		\begin{align} \label{eq:E_ul}
		\bm{\mathrm{E}}_{k}^{\mathrm{UL}} &=(\bm{I} - \bm{Q}_{k}^{-1/2}\bm{\beta}_{k}\bm{G}_{k}^{\mathrm{H}} \bm{H}_{k}^{\mathrm{H}} \bm{F}_{k}\bm{Q}_{k}^{1/2})\nonumber \\& \times(\bm{I} - \bm{Q}_{k}^{-1/2}\bm{\beta}_{k}\bm{G}_{k}^{\mathrm{H}} \bm{H}_{k}^{\mathrm{H}} \bm{F}_{k}\bm{Q}_{k}^{1/2})^{\mathrm{H}} \nonumber \\& \sum_{j \neq k} \bm{Q}_{k}^{-1/2}\bm{\beta}_{k}\bm{G}_{k}^{\mathrm{H}} \bm{H}_{j}^{\mathrm{H}} \bm{F}_{j}\bm{Q}_{j}\bm{F}_{j}^{\mathrm{H}} \bm{H}_{j} \bm{G}_{k}\bm{\beta}_{k}\bm{Q}_{k}^{-1/2} + \nonumber\\&+ \sigma_{n}^{2} \bm{Q}_{k}^{-1/2}\bm{\beta}_{k}\bm{G}_{k}^{\mathrm{H}}\bm{G}_{k}\bm{\beta}_{k}\bm{Q}_{k}^{-1/2}.	
		\end{align}
	\end{small}
	Then, based on  the  equivalent  UL/DL  channel  pair,  we  derive  a general  framework  for  joint  DL  MSE  design. First, in the \textit{UL channel}, we find the globally optimal powers $\bm{Q}$ according to the optimization problem under consideration; then, we update the UL receivers as MMSE filters (\ref{rup}) and we compute the associated per stream MSE values $ \varepsilon_{i}^{\mathrm{UL}}, \; \forall i$. Second, in the \textit{DL channel}, we find the DL power allocation $\bm{P}$ which achieves the same UL MSE values; and we update the DL receivers as MMSE filters (\ref{rdl}). Finally, we update $ \bm{W}_{k} $. 
	%
	%
	
	The matrix weighted per user MSE can be expressed as follows
	\begin{small}
		\begin{align}
		&\epsilon_{w,k}^{\mathrm{UL}} = \mathrm{tr} \big( \bm{W}_k \bm{E}_{k}^{\mathrm{UL}} \big) \\
		&= \mathrm{tr} \big( \bm{W}_{k} \big) +  \mathrm{tr} \big( \bm{W}_{k} \bm{Q}_{k}^{- 1/2} \bm{\beta}_{k} \bm{G}_{k}^{\mathrm{H}} \bm{H}_{k}^{\mathrm{H}} \bm{F}_{k}\bm{Q}_{k} \bm{F}_{k}^{\mathrm{H}} \bm{H}_{k} \bm{G}_{k} \bm{\beta}_{k}  \bm{Q}_{k}^{- 1/2} \big) \nonumber
		\\ &- 2 \mathrm{Re} \Big \{ \mathrm{tr}  \big( \bm{Q}_{k}^{ 1/2} \bm{W}_{k} \bm{Q}_{k}^{- 1/2} \bm{\beta}_{k} \bm{G}_{k}^{\mathrm{H}} \bm{H}_{k}^{\mathrm{H}} \bm{F}_{k}   \big) \Big \} \nonumber \\
		& + \sigma_{n}^{2} \mathrm{tr} \big( \bm{W}_{k} \bm{Q}_{k}^{- 1/2} \bm{\beta}_{k} \bm{G}_{k}^{\mathrm{H}} \bm{G}_{k} \bm{\beta}_{k} \bm{Q}_{k}^{- 1/2} \big)  \nonumber
		\\&+ \sum_{j \neq k} \mathrm{tr} \big( \bm{W}_{k} \bm{Q}_{k}^{- 1/2} \bm{\beta}_{k} \bm{G}_{k}^{\mathrm{H}} \bm{H}_{j}^{\mathrm{H}} \bm{F}_{j}\bm{Q}_{j} \bm{F}_{j}^{\mathrm{H}} \bm{H}_{j} \bm{G}_{k} \bm{\beta}_{k}  \bm{Q}_{k}^{- 1/2} \big), \forall k. 
		\nonumber
		\end{align}
	\end{small}
	We define $ \bm{Q}_k = \tilde{q}_k \bar{\bm{Q}}_k $ where $ \mathrm{tr}\big(\bar{\bm{Q}}_k\big) = 1$ and $ \tilde{q}_k$ is the individual power of the $k$th user. Then, the transmit covariance matrix $\bm{R}_k = \bm{F}_k \bm{Q}_k \bm{F}_{k}^{\mathrm{H}} $ can be written as $ \bm{R}_k = \tilde{q}_k \bar{\bm{R}}_k$ with $ \mathrm{tr}\big( \bar{\bm{R}}_k\big) = 1$. Thus, the matrix weighted MSE $ \epsilon_{w,k} $ becomes a function of $ \tilde{\bm{q}} = [ \tilde{q}_1, ..., \tilde{q}_K]^{T}$
	\begin{small}
		\begin{equation} \label{eq:eps}
		\epsilon_{w,k}^{\mathrm{UL}} = a_k + \tilde{q}_{k}^{-1} \sum_{j \neq k} \tilde{q}_j b_{kj} + \tilde{q}_{k}^{-1} c_k \sigma_{n}^{2}, \forall k 
		\vspace{-3mm}
		\end{equation}
	\end{small}
	where\vspace{-2mm}
	\begin{small}
		\begin{align}
		a_{k} &=  \mathrm{tr} \big( \bm{W}_{k} \big) + \mathrm{tr} \big( \bm{W}_{k} \bar{\bm{Q}}_{k}^{- 1/2} \bm{\beta}_{k} \bm{G}_{k}^{\mathrm{H}} \bm{H}_{k}^{\mathrm{H}} \bar{\bm{R}}_{k} \bm{H}_{k} \bm{G}_{k} \bm{\beta}_{k}  \bar{\bm{Q}}_{k}^{- 1/2} \big) \nonumber \\ &- 2 \mathrm{Re} \Big \{ \mathrm{tr}  \big(\bm{Q}_{k}^{ 1/2} \bm{W}_{k} \bm{Q}_{k}^{- 1/2} \bm{\beta}_{k} \bm{G}_{k}^{\mathrm{H}} \bm{H}_{k}^{\mathrm{H}} \bm{F}_{k}  \big) \Big \} , \nonumber\\
		b_{kj} &= \mathrm{tr} \big( \bm{W}_{k} \bar{\bm{Q}}_{k}^{- 1/2} \bm{\beta}_{k} \bm{G}_{k}^{\mathrm{H}} \bm{H}_{j}^{\mathrm{H}} \bar{\bm{R}}_{j} \bm{H}_{j} \bm{G}_{k} \bm{\beta}_{k}  \bar{\bm{Q}}_{k}^{- 1/2} \big) \nonumber
		\end{align}
	\end{small} and \begin{small}$c_{k}~ =~\mathrm{tr} \big( \bm{W}_{k} \bar{\bm{Q}}_{k}^{- 1/2} \bm{\beta}_{k} \bm{G}_{k}^{\mathrm{H}} \bm{G}_{k} \bm{\beta}_{k} \bar{\bm{Q}}_{k}^{- 1/2} \big)$.
	\end{small}
	
	Actually, problem (\ref{eq:primalup}) always has a global minimizer $\tilde{\bm{q}}^{\mathrm{opt}}$ characterized by the following equations:
	\begin{small}
		\begin{align}  \label{eq:del}
		\Delta^{\mathrm{UL}} & = \frac{\epsilon_{w,k}^{\mathrm{UL}}(\bm{q}^{\mathrm{opt}})}{ \xi_{k} } , \; \; \forall k,
		\\ \label{eq:qopt} \lVert \bm{q}^{\mathrm{opt}} \rVert_{1}  &= P_{\max} 
		\end{align}
	\end{small}
	where $\Delta^{\mathrm{UL}}$ is the minimum balanced matrix-weighted user MSE.
	
	We aim to form an eigensystem by combining (\ref{eq:del}) and (\ref{eq:qopt}). For that, we rewrite (\ref{eq:eps}) as
	\begin{small}
		\begin{equation} 
		\bm{\epsilon}_{w}^{\mathrm{UL}}\tilde{\bm{q}} = \bm{A} \tilde{\bm{q}} +  \sigma_{n}^{2}\bm{C}\bm{1}_{K}
		\end{equation}
	\end{small}
	where $\bm{\epsilon}_{w}^{\mathrm{UL}} = \text{diag} \{ \epsilon_{w,1}^{\mathrm{UL}}, \ldots ,\epsilon_{w,K}^{\mathrm{UL}} \} $, $\bm{C} = \text{diag} \big\{c_1, \ldots, c_K \big \} $ and
	\begin{small}
		\begin{equation}
		[\bm{A}]_{kj} = \begin{cases}
		b_{kj} ,
		& k \neq j \\
		a_k,			
		& k=j.
		\end{cases} \nonumber
		\end{equation} 
	\end{small}
	Now, we define $ \bm{\xi}= \text{diag} \big\{ [\xi_1 \ldots \xi_K] \big \} $ and multiply both sides by $\bm{\xi}^{-1} $ to have
	\begin{small}
		\begin{equation} \label{eq:h}
		\bm{\xi}^{-1}\bm{\epsilon}_{w}^{\mathrm{UL}}\tilde{\bm{q}} = \bm{\xi}^{-1}\bm{A} \tilde{\bm{q}} +  \sigma_{n}^{2}\bm{\xi}^{-1}\bm{C}\bm{1}_{K}.
		\end{equation}
	\end{small}
	From (\ref{eq:del}), we have $\bm{\xi}^{-1} \bm{\epsilon}_{w}^{\mathrm{UL}}(\tilde{\bm{q}}^{\mathrm{opt}}) = \Delta^{\mathrm{UL}} \bm{I}$. Thus, (\ref{eq:h}) becomes
	\begin{small}
		\begin{equation} \label{eq:hh}
		\Delta^{\mathrm{UL}}\tilde{\bm{q}} = \bm{\xi}^{-1}\bm{A} \tilde{\bm{q}}  +  \sigma_{n}^{2}\bm{\xi}^{-1}\bm{C}\bm{1}_{K}.
		\end{equation}
	\end{small}
	From (\ref{eq:qopt}), we can reparameterize $\tilde{\bm{q}} = \frac{P_{\max}}{\bm{1}_{K}^{\mathrm{T}} \bm{q}^{'}} \bm{q}^{'}$
	where $\bm{q}^{'}$ is unconstrained. This allows to rewrite (\ref{eq:hh}) as \cite{Negro11}
	\begin{equation} \label{eq:sys}
	\bm{\Lambda} \tilde{\bm{q}}^{'} = \Delta^{\mathrm{UL}} \tilde{\bm{q}}^{'} \, ,\;
	\bm{\Lambda} = \bm{\xi}^{-1}\bm{A}   +  \frac{\sigma_{n}^{2}}{P_{\max}}\bm{\xi}^{-1}\bm{C}\bm{1}_{K}\bm{1}_{K}^{\mathrm{T}}
	\end{equation}
	It can be observed that $\Delta^{\mathrm{UL}} $ is an eigenvalue of the non-negative extended coupling matrix $\bm{\Lambda}$. However, not all eigenvalues represent physically meaningful values. In particular, $\tilde{\bm{q}}^{\mathrm{opt}} > 0$ and $\Delta^{\mathrm{UL}} >0$ must be fulfilled.
	
	It is known that for any non-negative irreducible real matrix $\bm{X}$ with spectral radius $\rho(\bm{X})$, there exists a unique vector $\bm{q}>0$ and $\lambda_{\max}(\bm{X}) = \rho(\bm{X})$ such that $\bm{X} q=  \lambda_{\max}(\bm{X}) \bm{q}$. The uniqueness of $\lambda_{\max}(\bm{\Lambda}) $ also follows from immediately from the function $\Delta^{\mathrm{UL}} (P_{\max})$ being strictly monotonically decreasing in $P_{\max}$. This rules out the existence of two different balanced levels with the same sum power.  Hence, the balanced level is given by\vspace{-2mm}
	\begin{small}
		\begin{equation}
		\Delta^{\mathrm{UL, \; opt}} = \lambda_{\max}(\bm{\Lambda}).
		\vspace{-2mm}
		\end{equation}
	\end{small}
	Therefore, the optimal power allocation $\tilde{\bm{q}}^{'}$ is the principal eigenvector of the matrix $\bm{\Lambda}$
	in (\ref{eq:sys}). As noted in \cite{Boche4}, we have in fact\vspace{-2mm}
	\begin{equation}
	\lambda_{\max}(\bm{\Lambda}) = \min_{\tilde{\bm{p}}}\max_{\tilde{\bm{q}}}
	\frac{\tilde{\bm{p}}^H\bm{\Lambda}\tilde{\bm{q}}}{\tilde{\bm{p}}^H\tilde{\bm{q}}}
	= \max_{\tilde{\bm{p}}}\min_{\tilde{\bm{q}}}
	\frac{\tilde{\bm{p}}^H\bm{\Lambda}\tilde{\bm{q}}}{\tilde{\bm{p}}^H\tilde{\bm{q}}}
	\label{eq:pq}
	\end{equation}
	where in \cite{Boche4} $\tilde{\bm{p}}$ was said to have no particular meaning but actually can be shown to relate to the DL powers.
	So, the proposed algorithm provides in the inner loop an alternating optimization of (\ref{eq:pq}) w.r.t.
	$\tilde{\bm{p}}$, $\tilde{\bm{q}}$, $\bm{F}$, $\bm{G}$ \cite{Boche4}, \cite{shi8}.
	If we take for $\tilde{\bm{p}}$ the $K$ standard basis vectors, then we get\vspace{-2mm}
	\begin{equation}
	\lambda_{\max}(\bm{\Lambda}) = \min_{\tilde{\bm{q}}}\max_k \frac{\big(\bm{\Lambda}\tilde{\bm{q}}\big)_k}{\tilde{\bm{q}}_k}
	\vspace{-2mm}
	\label{eq:pq2}
	\end{equation}
	which from (\ref{eq:del}), (\ref{eq:h}), (\ref{eq:sys}) can be seen to be exactly the WMSE balancing problem we want to solve.
	
	\begin{table}
		\caption{Pseudo code of the proposed algorithm} \label{tab:alg}\vspace{-2mm}
		\begin{tabular}{lp{7.8cm}}
			\hline
			& \vspace{-2mm}\\
			1. & initialize: $\bm{F}_{k}^{\mathrm{H}(0,0)} = (\bm{I}_{d_{k}}\colon \bm{0})$, $\bar{\bm{Q}}^{(0,0)} = \frac{P_{\max}}{N_d}\bm{I}$, $m =n = 0$ and  $n_{\max}, m_{\max}$ and fix $r_{k}^{\circ (0)}$\\
			2. & compute UL receive filter $\bm{G}^{(0,0)}$ and $\bm{\beta}^{(0,0)}$ with (\ref{rup})\\
			3. & set $\bm{W}_{k}^{(0)} = \bm{I}$ and  $\xi_{k}^{(0)} = d_k $  \\
			4. & find optimal user power allocation $\bm{\tilde{q}}^{(0,0)}$ by solving (\ref{eq:sys}) and compute $\bm{Q}_{k}^{(0,0)} = \tilde{q}_{k}^{(0,0)} \bm{\bar{Q}}_{k}^{(0,0)}$\\
			5. &\textbf{repeat}\vspace{-7mm}\\
			&
			\begin{itemize}
				\item[5.1]  \textbf{repeat} 
				\item[] $n \leftarrow n +1$
				\item[] \textit{UL channel:}
				\item[$\bullet$] update $ \bm{G}^{(n,m-1)} $ and $\bm{\beta}^{(tmp,tmp)} $ with (\ref{rup})
				\item[$\bullet$] compute the MSE values $ \bm{\varepsilon}^{\mathrm{UL}, (n)} $ with (\ref{eq:upmse})
				\item[] \textit{DL channel:}
				\item[$\bullet$] compute $\bm{P}^{(n,m-1)}$ with (\ref{eq:p})
				\item[$\bullet$] update $\bm{F}^{(n,m-1)}$ and $\bm{\beta}^{(tmp,tmp)} $ with (\ref{rdl})
				\item[$\bullet$] compute the MSE values $ \bm{\varepsilon}^{\mathrm{DL} (n)} $ with (\ref{eq:dlmse})
				\item[] \textit{UL channel:}
				\item[$\bullet$] compute $\bm{Q}^{(tmp, tmp)}$ with (\ref{eq:q}) and $\bar{\bm{Q}}_{k}^{(n,m-1)} = \bm{Q}_{k}^{(tmp, tmp)}/ \mathrm{tr} \big(\bm{Q}_{k}^{(tmp, tmp)}\big) $
				\item[$\bullet$] find optimal user power allocation $\bm{\tilde{q}}^{(n,m-1)}$ by solving (\ref{eq:sys}) and compute $\bm{Q}_{k}^{(n,m-1)} = \tilde{q}_{k}^{(n,m-1)} \bm{\bar{Q}}_{k}^{(n, m-1)}$
				\item[5.2] \textbf{until} required accuracy is reached or $n \geq n_{\max}$
				\item[5.3]  $ m \leftarrow m +1$
				\item[5.4] update $\bm{W}_{k}^{(m) }= (\bm{\mathrm{E}}_{k}^{\mathrm{UL} (m)})^{-1} $, $r_k^{(m)} = \ln\det (\bm{W}_{k}^{(m) })$,
				$t = \min_{k} \frac{r_{k}^{(m) } }{r_{k}^{\circ (m-1)}} $, 
				$r_{k}^{\circ (m)} = t\, r_{k}^{\circ (m-1)} $,  and $\xi_{k}^{(m)} = d_k + r_k^{(m)} - r_{k}^{\circ (m)}$
				\item[5.5] do $n \leftarrow 0 $ and set $(.)^{(n_{\max} , m-1)} \rightarrow (.)^{(0,m)} $ in order to re-enter the inner loop
				\vspace{-3mm}
			\end{itemize} \\
			6. & \textbf{until} required accuracy is reached or $m \geq m_{\max} $ \\
			\hline 
			\vspace{-6mm}
		\end{tabular}
	\end{table}
	\vspace{-2mm}
	
	\section{Algorithmic Solution and Simulations}
	\vspace{-1mm}
	\label{sec:algo}
	
	\subsection{Algorithm}\vspace{-2mm}
	
	The proposed optimization framework is summarized hereafter in Table~\ref{tab:alg}. Superscripts $(.)^{(n)}$ and $(.)^{(tmp)}$ denote the $n^{\mbox{th}}$ iteration and a temporary value, respectively. 
	This algorithm is based on a double loop. The inner loop solves the WMSE balancing problem in (\ref{eq:primal})
	whereas the outer loop iteratively transforms the WMSE balancing problem into the original rate balancing problem in (\ref{eq:main}).
	\vspace{-2mm}
	%
	
	\subsection{Proof of Convergence}\vspace{-2mm}
	
	In case the rate weights $r^{\circ}_k$ would not satisfy $r_k \geq r^{\circ}_k$, this issue will be rectified by the scale factor $t$ after one iteration (of the outer loop). 
	It can be shown that $t = \min_{k} \frac{r_{k}^{(m) } }{r_{k}^{\circ (m-1)}} \geq 1$. 
	By contradiction, if this was not the case, it can be shown to lead to
	$\frac{\mathrm{tr} \big( \bm{W}_{k}^{(m-1)}  \bm{\mathrm{E}}_{k}^{(m)}  \big) }{\xi_k^{(m-1)}} > 1, \;\forall k$
	and hence $\Delta^{(m)} > 1$.
	But we have 
	\begin{equation}\!\!\!\!
	\begin{array}{r@{\!}c@{\!}l}
	\Delta^{(m)}\,  & = & \,
	\frac{\mathrm{tr} \big( \bm{W}_{k}^{(m-1)}  \bm{\mathrm{E}}_{k}^{(m)}  \big) }{\xi_k^{(m-1)}},\; \forall k ,
	=\max_k \frac{\mathrm{tr} \big( \bm{W}_{k}^{(m-1)}  \bm{\mathrm{E}}_{k}^{(m)}  \big) }{\xi_k^{(m-1)}}\\
	& \stackrel{(a)}{<} & \,\max_k \frac{\mathrm{tr} \big( \bm{W}_{k}^{(m-1)}  \bm{\mathrm{E}}_{k}^{(m-1)}  \big) }{\xi_k^{(m-1)}}
	=\max_k \frac{d_k}{\xi_k^{(m-1)}} \, \stackrel{(b)}{<} 1\, .
	\end{array}
	\label{eq:DeltaUB}
	\end{equation}
	Let $\bm{\mathrm{E}} = \{\bm{\mathrm{E}}_k, k=1,...,K\}$ and \\
	$f^{(m)}(\bm{\mathrm{E}}) = \max_k \frac{\mathrm{tr} \big( \bm{W}_{k}^{(m-1)}  \bm{\mathrm{E}}_{k} \big) }{\xi_k^{(m-1)}}$.
	Then $(a)$ is due to the fact that the algorithm in fact performs
	alternating minimization of $f^{(m)}(\bm{\mathrm{E}})$ w.r.t.\  $\bm{G}$, $\bm{F}$, $\tilde{\bm{q}}$ and hence will lead to 
	$f^{(m)}(\bm{\mathrm{E}}^{(m)}) < f^{(m)}(\bm{\mathrm{E}}^{(m-1)}) $. 
	On the other hand, $(b)$ is due to 
	$\xi_k^{(m-1)} = d_k + r_k^{(m-1)} - r_k^{\circ (m-1)} > d_k $, for $m\geq3$.
	
	Hence, $t\geq 1$.  Of course, during the convergence $t>1$. 
	The increasing rate targets $r_k^{\circ (m)}$ constantly catch up with the increasing rates $r_k^{(m)}$.
	Now, the rates are upper bounded by the single user MIMO rates (using all power), and hence the rates will converge
	and the sequence $t$ will converge to 1. That means that for at least one user $k$, $r_k^{(\infty)} = r_k^{\circ (\infty)}$.
	The question is whether this will be the case for all users, as is required for rate balancing.
	Now, the WMSE balancing leads at every outer iteration $m$ to 
	$\frac{\mathrm{tr} \big( \bm{W}_{k}^{(m-1)}  \bm{\mathrm{E}}_{k}^{(m)}  \big) }{\xi_k^{(m-1)}} = \Delta^{(m)}, \forall k$.
	At convergence, this becomes $\frac{d_k }{\xi_k^{(\infty)}} = \Delta^{(\infty)}$ where 
	$\xi_k^{(\infty)} = d_k + r_k^{(\infty)} - r_k^{\circ (\infty)}$. Hence, if we have convergence because for one user $k_{\infty}$ we arrive at
	$r_{k_{\infty}}^{(\infty)} =r_{k_{\infty}}^{\circ (\infty)}$, then this implies $\Delta^{(\infty)} = 1$ which implies 
	$r_{k}^{(\infty)} =r_{k}^{\circ (\infty)}, \forall k$. Hence, the rates will be maximized and balanced.
	\vspace{-2mm}
	
	\begin{remark}
		In fact, the algorithm also converges with $n_{\max} = 1$, i.e., with only a single loop. 
	\end{remark}
	\vspace{-3mm}

	\subsection{Simulation results}
	
	In this section, we numerically illustrate the performance of the proposed algorithm. The simulations are obtained under a channel modeled as follows : $\bm{H}_{k}^{\mathrm{H}} = \bm{B}_{k} \bm{\mathcal{U}}_{k} \bm{A}_{k} $ where $\bm{B}_{k}, \bm{A}_{k}  $ are of dimensions $(M\times N_k)$ and $(N_k\times N_k)$ respectively, and
	have i.i.d. elements distributed as $\mathcal{CN}(0,1)$; $\bm{\mathcal{U}}_{k} = \mu \bm{U}_{k} $, with the normalization parameter $\mu = (\text{trace} \big ( \bm{U}_{k}) \big )^{-1/2}$ and $ \bm{U}_{k} = \text{diag} \big \{ 1, \alpha, \alpha^{2}, \ldots, \alpha^{N_k -1} \big\}$ ($\alpha \in \mathbb{R}$ being a scalar parameter). This model allows to control the rank profile of the MIMO channels. For all simulations, we fix $\alpha = 0.3$ and take 1000 channel realisations and $n_{\max}=20$. The algorithm converges after 4-5 (or 13-15) iterations of $m$ at SNR = $ \frac{P_{\max}}{\sigma_{n}^{2}} =$10dB (or 30dB). 
	
	%
	
\begin{figure}
		\vspace{-3mm}
		\hspace{-7mm}
		\includegraphics[width=10.5cm,height=6.5cm]{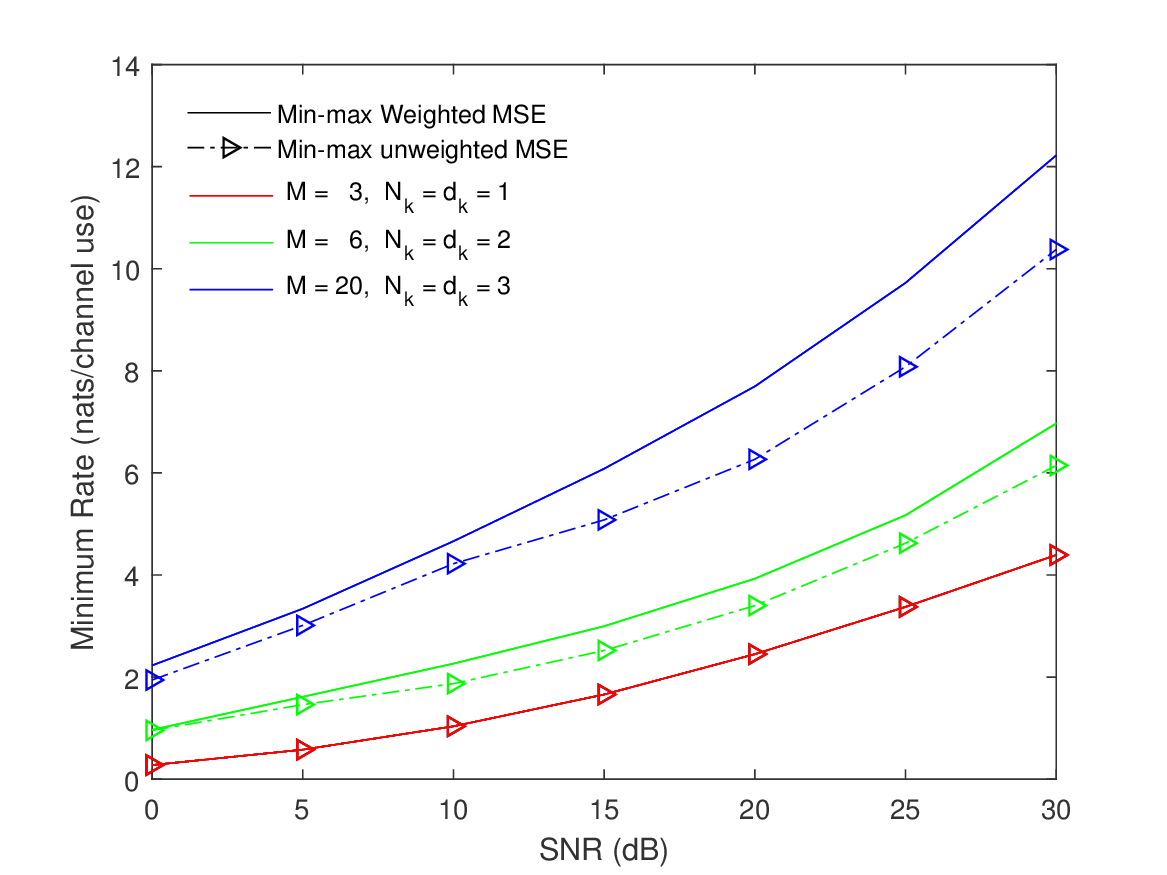}
		\vspace{-9mm}
		\caption{Minimum rate in the system VS SNR: $K=3$.}
		\label{fig:minrate}
	\end{figure}

	Figure \ref{fig:minrate} plots the minimum achieved per user rate using \textit{i)} our max-min user rate approach with equal user priorities and \textit{ii)} the user MSE balancing approach \cite{shi8}, as a function of the Signal to Noise Ratio (SNR). We observe that our approach outperforms significantly the unweighted MSE balancing optimization, and the gap gets larger with more streams. Note that we observe the same behavior with the classical i.i.d. channel $\bm{H}_{k}^{\mathrm{H}} = \bm{B}_{k}$	, but with a smaller gap (e.g., for 15dB, $\frac{\min_k r_k(\text{weighted-MSE})}{\min_k r_k(\text{unweighted-MSE})} $ = 1.05 instead of 1.18 with $M = 6, N_k = d_k =2 $ in Figure \ref{fig:minrate}).
	
	\begin{figure}
		\vspace{-4mm}
		\hspace{-9mm}
		\includegraphics[width=10.75cm,height=6.5cm]{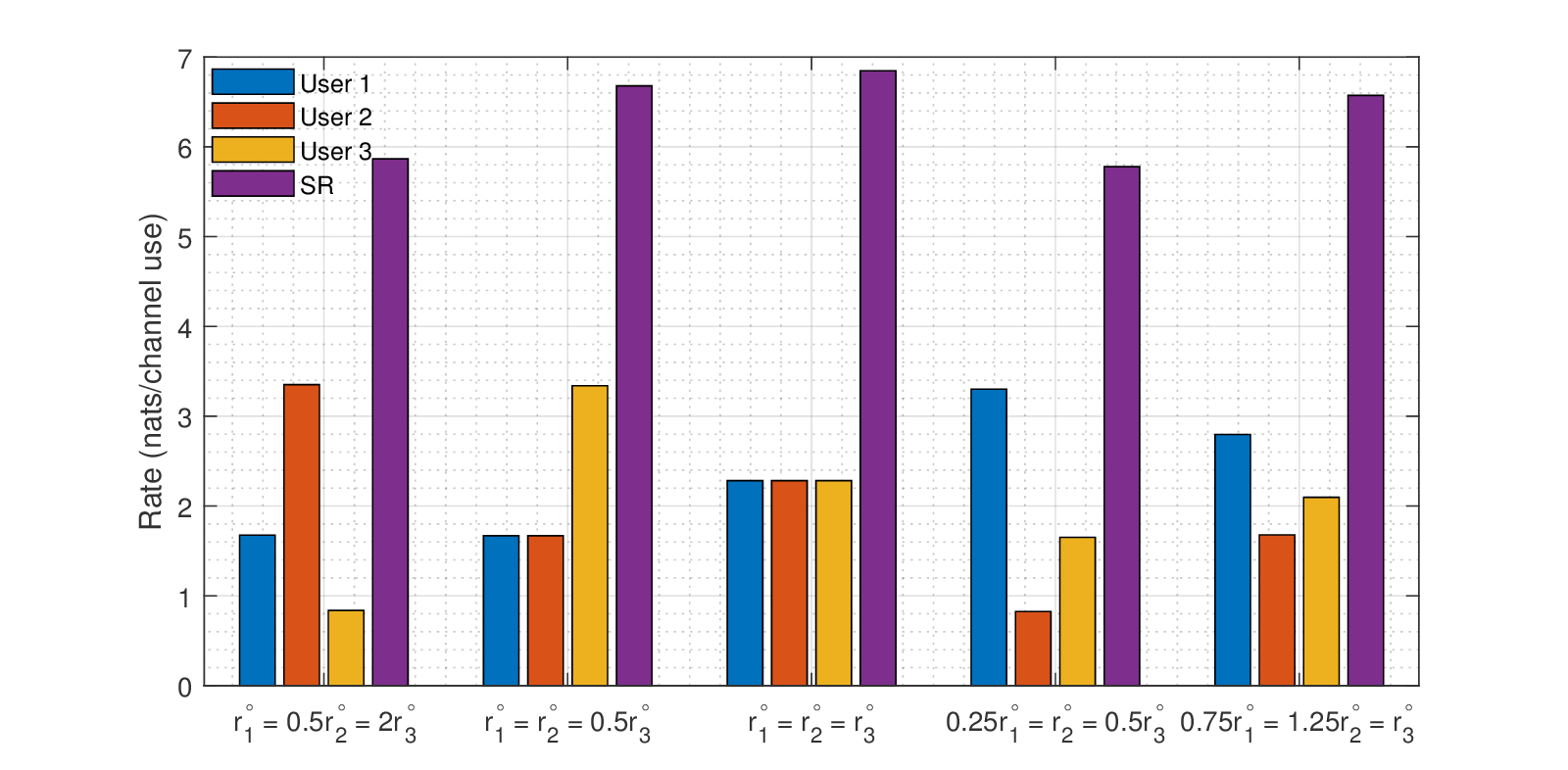}
		\vspace{-10mm}
		\caption{Rate distribution among users: $K=3$, SNR= $10$ dB, $M= 6, N_k = d_k = 2$.}
		\vspace{-7mm}
		\label{fig:rate}
	\end{figure}
	In Figure \ref{fig:rate}, we illustrate how rate is distributed among users according to their priorities represented by the rate targets $r^{\circ}_{k}$. We can see that, using the min-max weighted MSE approach, the rate is equally distributed between the users with equal user priorities, i.e.,  $r^{\circ}_{k} =r^{\circ}_{1} \; \forall k$, whereas with different user priorities, the rate differs from one user to another accordingly. Furthermore, the Sum Rate (SR) reaches its maximum when user priorities are equal, as the channel statistics are identical for each user.
	\vspace{-2mm}
	
	\section{Conclusions}
	\label{sec:Conclusions}
	\vspace{-1mm}
	
	In this work, we addressed the multiple streams per user case (MIMO links) for which we considered user rate balancing, not stream rate balancing. Actually, we optimized the rate distribution over the streams of a user, within the rate balancing of the users. In this regard, we proposed an iterative algorithm to balance the rate between the users in a MIMO system. The latter was derived by transforming the max-min rate optimization problem into a min-max weighted MSE optimization problem to enable MSE duality. We also provided numerical comparisons between the proposed weighted rate balancing approach and unweighted MSE balancing.
	\vspace{-1mm}
	
	\section*{Acknowledgments}
	\vspace{-1mm}
	
	This work has been performed in the framework of the Horizon 2020 project ONE5G (ICT-760809) receiving funds from the European Union. The authors would like to acknowledge the contributions of their colleagues in the project, although the views expressed in this contribution are those of the authors and do not necessarily represent the project.
	EURECOM's research is partially supported by its industrial members:
	ORANGE, BMW, Sy\-man\-tec, SAP, Monaco Telecom, iABG,  and by the projects DUPLEX (French ANR)
	and MASS-START (French FUI).
	
	
	\bibliographystyle{IEEEtran}
	\vspace{-2mm}

			\bibliography{ref}

\end{document}